\documentclass[useAMS,usenatbib]{mn2e}
\usepackage{times,epsfig,graphics,graphicx,multicol,amssymb,paralist,mn2e-breakabs,amsmath}
\usepackage{natbib}

\def\lir{L$_{IR}$}
\def\halpha{H$\alpha$}
\def\SFRSFR{SFR(L$_{IR}$)-SFR(H$\alpha$)}
\def\ha{H$\alpha$}
\def\hb{H$\beta$}

\begin{document}

\title[Comparison of star formation rate from \ha~and infrared luminosity]{Comparison of star formation rates from \ha~and infrared luminosity as seen by Herschel\thanks{\textit{Herschel} is an ESA space observatory with science instruments
provided by European-led Principal Investigator consortia and with important participation from NASA.}} 
\author[H. Dom{\'i}nguez S{\'a}nchez et al.]{H. Dom{\'i}nguez S{\'a}nchez$^{1,2}$, M.Mignoli$^1$, F.~Pozzi$^3$, F.~Calura$^1$, A.~Cimatti$^3$, C.~Gruppioni$^1$, 
 \newauthor
J. Cepa $^2$, M. S{\'a}nchez Portal$^4$, G.~Zamorani $^1$, 
S.~Berta$^{5}$, D.~Elbaz$^6$, E.~Le Floc'h$^{7}$,
\newauthor
 G.~L.~Granato$^8$, D.~Lutz$^{5}$,  R.~Maiolino$^9$, F.~Matteucci$^{10}$, P.~Nair $^{1, 11}$, R.~Nordon$^{5}$, L.~Pozzetti$^{1}$, 
\newauthor
 L.~Silva$^{8}$, J. Silverman$^{12}$,  S.~Wuyts$^{5}$, 
 C.~M.~Carollo$^{13}$,
 T.~Contini$^{14}$,
 J.-P.~Kneib$^{15}$,
\newauthor
 O.~Le~F\`evre$^{15}$,
 S.~J.~Lilly$^{13}$,
 V.~Mainieri$^{16}$,
 A.~Renzini$^{17}$,
 M.~Scodeggio$^{18}$,
 S.~Bardelli$^1$,
\newauthor
 M.~Bolzonella$^1$,
 A.~Bongiorno$^{5}$,
 K.~Caputi$^{19}$,
 G.~Coppa$^{1,3}$,
 O.~Cucciati$^{8}$,
 S.~de~la~Torre$^{18,20}$,
\newauthor
 L.~de~Ravel$^{15,19}$,
 P.~Franzetti$^{18}$,
 B.~Garilli$^{18}$,
 A.~Iovino$^{20}$,
 P.~Kampczyk$^{13}$,
 C.~Knobel$^{13}$,
\newauthor
 K.~Kova\v{c}$^{13}$,
 F.~Lamareille$^{14}$,
  J.-F.~Le~Borgne$^{14}$,
 V.~Le Brun$^{15}$,
 C.~Maier$^{13}$,
 B.~Magnelli$^{5}$,
\newauthor
 R.~Pell\'o$^{14}$,
 Y.~Peng$^{13}$,
E.~Perez-Montero$^{14,21}$,
 E.~Ricciardelli$^{17,2}$,
 L.~Riguccini$^{7}$
 M.~Tanaka$^{8,17}$,
\newauthor
 L.~A.~M.~Tasca$^{15,18}$,
 L.~Tresse$^{15}$,
 D.~Vergani$^{1}$, 
 E.~Zucca$^{1}$
\\
$^1$INAF-Osservatorio Astronomico di Bologna, via Ranzani 1, I-40127 Bologna, Italy; E-mail: helena.dominguez@oabo.inaf.it\\
$^2$Instituto de Astrof{\'i}sica de Canarias, 38205 La Laguna, Spain\\
$^3$Alma Mater Studiorum Universit\`a di Bologna, Dipartimento di Astronomia, via Ranzani 1, I-40127 Bologna, Italy\\
$^4$European Space Astronomy Centre ESAC/ESA, Madrid, Spain\\
$^{5}$Max-Planck-Institut f\"ur Extraterrestrische Physik, Giessenbachstra{\ss}e,85748 Garching bei M\"unchen, Germany\\ 
$^{6}$CEA Saclay, Service d'Astrophysique, Orme des Merisiers, Bat.709, F-91191 Gif-sur-Yvette Cedex, France\\
$^{7}$Laboratoire AIM, CEA/DSM-CNRS-Universite Paris Diderot, IRFU/Service d'Astrophysique, CEA-Saclay, 91191 Gif-sur-Yvette Cedex, France\\
$^{8}$INAF - Osservatorio Astronomico di Trieste, via G. B. Tiepolo 11, 34143 Trieste, Italy\\ 
$^9$INAF - Osservatorio Astronomico di Roma, via di Frascati 33, 00040 Monte Porzio, Italy\\
$^{10}$Dipartimento di Fisica - Sezione di Astronomia - Università degli Studi di Trieste - Piazzale Europa, 1, I-34127 Trieste,  Italy\\
$^{11}$Space Telescope Science Institute, 3700 San Martin Drive, Baltimore, MD, USA\\
$^{12}$Institute for the Physics and Mathematics of the Universe (IPMU), University of Tokyo, Kashiwanoha 5-1-5, Kashiwa-shi, Chiba 277-8568, Japan\\ 
$^{13}$ETH Zurich, Institute of Astronomy, Wolfgang-Pauli-Stra{\ss}e 27, 8093 Zurich, Switzerland\\ 
$^{14}$Laboratoire d'Astrophysique de Toulouse-Tarbes, Universit\'{e} de Toulouse, CNRS, 14 avenue Edouard Belin, 31400 Toulouse, France\\ 
$^{15}$Laboratoire d'Astrophysique de Marseille, Universit\'{e} d'Aix-Marseille, CNRS, 38 rue Fr\'{e}d\'{e}ric Joliot-Curie, 13388 Marseille Cedex 13, France\\ 
$^{16}$European Southern Observatory, Karl-Schwarzschild-Stra{\ss}e 2, 85748 Garching bei M\"unchen, Germany\\ 
$^{17}$INAF - Osservatorio Astronomico di Padova, vicolo dell'Osservatorio 5, 35122 Padova, Italy\\ 
$^{18}$INAF - IASF Milano, via Bassini 15, 20133 Milano, Italy\\ 
$^{19}$Kapteyn Astronomical Institute, University of Groningen, P.O. Box 800, 9700 AV Groningen, The Netherlands\\ 
$^{20}$INAF - Osservatorio Astronomico di Brera, via Brera 28, 20121 Milano, Italy\\ 
$^{21}$Instituto de Astrofisica de Andalucia, CSIC, Apdo. 3004, 18080 Granada, Spain \\
}

\date{} 
\maketitle 
 

\begin{abstract}

We empirically test the relation between the SFR(\lir) derived from the infrared luminosity, \lir,  and the SFR(\ha) derived from the \ha~emission line luminosity using simple conversion relations. We use a sample of 474 galaxies  at $z=0.06-0.46$ with both \ha~detection (from 20k zCOSMOS survey) and new far-IR Herschel data (100 and 160 $\mu$m). We derive  SFR(\ha) from the H${\alpha}$ extinction corrected emission line luminosity. We find a very clear trend between $E(B-V)$ and \lir~that allows to estimate extinction values for each galaxy even if the \hb~emission line measurement is not reliable. We calculate the L$_{IR}$ by integrating from 8 up to 1000 $\mu$m the SED that is best fitting our data. We compare the SFR(\halpha) with the  SFR(L$_{IR}$). We find a very good agreement between the two SFR estimates, with a slope of $m=1.01 \pm 0.03$ in the log SFR(L$_{IR}$) vs log SFR(H${\alpha}$) diagram, a normalization constant of $a=-0.08 \pm 0.03$ and a dispersion of $\sigma=0.28$ dex. We study the effect of some intrinsic properties of the galaxies in the \SFRSFR~relation, such as the redshift, the mass, the SSFR or the metallicity. The metallicity is the parameter that affects most the SFR comparison. The mean ratio of the two SFR estimators  log[SFR(\lir)/SFR(\halpha)] varies by  $\sim 0.6$ dex from  metal-poor to metal-rich galaxies ($8.1 <log (O/H) + 12 < 9.2$). This effect is consistent with the prediction of a theoretical model for the dust evolution in spiral galaxies. Considering different morphological types, we find a very good agreement between the two SFR indicators for the Sa, Sb and Sc morphologically classified galaxies, both in slope and normalization. For the Sd, irregular sample (Sd/Irr), the formal best-fit slope becomes much steeper ($m=1.62 \pm 0.43$), but it is still consistent  with 1 at the $1.5\sigma$ level, because of the reduced statistics of this sub-sample.
%
%


\end{abstract}

\section{Introduction}

The star formation rate (SFR) is one of the most important parameters to study galaxy evolution, as it accounts for the number of stars being formed in a galaxy per unit time. Many authors have studied the evolution of the star formation rate density with time, showing that the peak of the star formation took place at $z\sim1-2$, then slowing down at later times (see the data collection from different surveys by \citealt{Hopkins2006} and recent Herschel results from \citealt{Gruppioni2010}). There is also agreement in the fact that SFR directly correlates with the galaxy properties such as gas content, morphology or mass. Recent papers such as \cite{Elbaz2011}, \cite{Wuyts2011b}, \cite{Daddi2010}, \cite{Rodighiero2010} or \cite{Noeske2009} have investigated in detail the existence of a main sequence in the SFR-mass relation. Therefore, it is fundamental to have a rigorous way to measure SFR when studying galaxy evolution.

A great effort has been made over the past years to derive SFR indicators from luminosities at different wavelengths, spanning from the UV, where the recently formed massive stars emit the bulk of their energy; to the infrared, where the dust-reprocessed light from those stars emerges; to the radio, which traces supernova activity; to the X-ray luminosity, tracing X-ray binary emission (e.g., \citealt{Kennicutt1998}, \citealt{Yun2001}, \citealt{Kewley2002}, \citealt{Ranalli2003}, \citealt{Calzetti2005}, \citealt{Calzetti2007}, \citealt{Schmitt2006}, \citealt{Alonso-Herrero2006}, \citealt{Rosa-Gonzalez2007}, \citealt{Calzetti2009}, \citealt{Calzetti2010}). Recent works combine optical and infrared observations to derive attenuation-corrected  H${\alpha}$  and UV continuum luminosities of galaxies by combining these fluxes with various components of IR tracers (e.g. \citealt{Gordon2000}, \citealt{Inoue2001}, \citealt{Hirashita2001}, \citealt{Bell2003}, \citealt{Hirashita2003}, \citealt{Iglesias-Paramo2006}, \citealt{Cortese2008}, \citealt{Kennicutt2009}, \citealt{Wuyts2011a}).

The SFR derived from the \ha~emission line is calibrated on the physical basis of photoionization. The nebular lines effectively re-emit the integrated stellar luminosity of galaxies short-ward the Lyman limit, so they provide a direct, sensitive probe of the young massive population. Only stars with masses  $> 10 $M$_{\odot}$ and lifetimes   $<$ 20 Myr contribute significantly to the integrated ionizing flux, so the emission lines provide a nearly instantaneous measure of the SFR, independent of the previous star formation history. However, a large part of the light emitted by young stars, which mostly reside within or behind clouds of gas and dust, is absorbed by dust and then re-emitted at longer wavelengths. To properly calculate the SFR from \ha~emission line an extinction correction must be applied to take this effect into account.

The absorbed light from these young stars, tracers of the SFR, is re-emitted at IR wavelengths. Therefore, the SFR can also be derived from the IR emission of the galaxies. One of the most commonly used methods to derive the SFR from the IR emission is to apply the \cite{Kennicutt1998} [K98 hereafter] relation, which links the SFR to the IR luminosity (L$_{IR}$). However, this is a theoretical relation based on the starburst synthesis models of \cite{Leitherer-Heckman1995} obtained for a continuous burst, solar abundances and a \cite{Salpeter1955} Initial Mass Function (IMF), assuming that young stars dominate the radiation field throughout the UV-visible and that the FIR luminosity measures the bolometric luminosity of the starburst. This physical situation holds in dense circumnuclear starbursts that power many IR-luminous galaxies. In the disk of normal galaxies or early type galaxies, the situation is much more complex as dust heating from the radiation field of older stars may be very important. Strictly speaking, as remarked by K98, the relation above applies only to starburst with ages less than $10^8$ years while, for other cases, it would be probably better to rely on an empirical calibration of SFR/L$_{IR}$. 

The advent of the Herschel Space Telescope, which  samples the critical far-infrared peak of nearby galaxies, is an excellent opportunity to empirically derive the relation between the observed L$_{IR}$ and the SFR without associated uncertainties due to extrapolation of the IR peak. In this paper we analyze a far-IR  (100 or 160 $\mu$m) selected sample in the COSMOS field with associated multi-band photometry and 20k zCOSMOS optical spectroscopy at $z=0.1-0.46$. We determine the SFR estimates from the  H${\alpha}$ line luminosity corrected for extinction, while we obtain the L$_{IR}$  by integrating from 8 to 1000 $\mu$m the best fitting SED of our galaxies. We test if the SFR derived from the \ha~emission and the SFR derived from the L$_{IR}$ are consistent at different redshifts and physical properties of the galaxies (mass,  metallicity, specific star formation rate or morphological type).

This paper is organized as follows: Section 2 introduces our sample of galaxies. Sections 3 and 4 explain the method used to derive the fundamental physical parameters of our sample (L$_{IR}$, stellar masses, SFR from dust extinction corrected H${\alpha}$ emission line). Section 5 presents the comparison of the two SFR indicators, while in Section 6 we discuss the effect of different galaxy properties on the comparison of  the two SFR estimates. Finally, in Section 7 we summarize our conclusions. 

Throughout this paper we use a standard cosmology ($\Omega_{m}=0.3,\Omega_{\Lambda}=0.7$), with $H_{0}=70$  km s$^{-1}$ Mpc$ ^{-1}$. The stellar masses are given in units of solar masses (M$_{\odot}$), and both the SFRs and the stellar masses assume a Salpeter IMF.

\section{Data}

We study the SFR derived from the L$_{IR}$ making use of the recent Herschel data in the COSMOS field. The Herschel data includes PACS \citep{Poglitsch2010} data from the PEP survey (PACS Evolutionary Probe, \citealt{Lutz2011}). We use the PACS blind catalog v2.1 (containing 7313 sources) down to $3\sigma$, corresponding to  $\sim ~5$ and  $10.2$ mJy
at 100 and 160 $\mu$m, respectively. This catalog was associated using the likelihood ratio technique \citep{Sutherland1992} to the deep  24 $\mu$m catalog from Le Floch et al. (private communication S$_{lim}$(24 $\mu$m)=80 $\mu$Jy; \citealt{LeFloch2009}, \citealt{Sanders2007}) and to the multi-band 3.6 $\mu$m selected catalog from \cite{Ilbert2010} (S$_{lim}$(3.6 $\mu$m)=1.2 $\mu$Jy), including \textit{$NUV$, $u^{\ast}$, $ B_{J}$, $ g^{+}$, $ V_{J}$,$ r^{+}$, $ i^{+}$, $z^{+}$, $J$,$K_{s}$, 3.6, 4.5, 5.8, 8.0} and 24 $\mu$m from GALEX, Subaru, IRAC and Spitzer legacy surveys, respectively.
We also need the spectroscopic information to accurately measure the SFR from the H${\alpha}$ emission line. We have 1654 sources ($\sim$ 23 $\%$ of PEP catalog) with spectroscopic information from the zCOSMOS 20k spectroscopic survey \citep{Lilly2009}, 717 of them with z $<$ 0.46. This cut in redshift is necessary to properly observe the H${\alpha}$ emission line (6562.8 $\AA$). We did not consider sources with no accurate redshifts (flag z $\leq$ 2.1, \citealt{Lilly2009}, $\sim 5\%$), neither sources with S/N $<$ 3 in the  \ha~flux ($\sim 25\%$). We did not consider galaxies with very high values of aperture correction ($>$ 16; 4 galaxies) to exclude significant uncertainties associated to this correction.

To avoid AGN contamination we have cleaned our sample of possible AGN candidates. Type-1 AGN's are easily recognizable by their broad permitted emission lines and have been excluded originally from the zCOSMOS galaxy sample. We have eliminated sources classified as Seyfert2 or LINERS using the BPT diagnostic diagrams from \cite{Bongiorno2010}, as well as XMM-Newton \citep{Hasinger2007} and Chandra detected sources \citep{Elvis2009}. We have also performed a SED fitting from the NUV to the PEP bands to our sources using the \cite{Polletta2007} templates which include  different SEDs, from elliptical and spiral normal galaxies, to starburst, composite, Seyfert1 and Seyfert2. We also used the modified templates by \cite{Gruppioni2010}, which include a higher IR bump to better fit the observed far-IR data. We have carefully inspected the resulting SED fits and we have eliminated those sources with a typical SED of a Seyfert. This procedure allows us to identify AGN sources which could be missed by the other methods (e.g., galaxies with too low S/N to be used in the BPT diagrams). We identify 47 AGN candidates ($\sim6\%$) and eliminate them.
The final sample consists of 474  PEP detected H${\alpha}$-emitting galaxies with $0.06 < z < 0.46$.  In Table \ref{prop} we summarize some of the main statistical properties of our sample (z, stellar mass, SFR, SSFR, metallicty and \ha, \hb~ equivalent widths)\footnote{PEP images and catalogs will be released at http://www.mpe.mpg.de/ir/Research/PEP/public$\_$data$\_$releases.php}.

\begin{table*}
\begin{tabular}{|l|c|c|c|c|c|}
\hline
 & Min & 1$^{st}$ quartile & Median & 3$^{rd}$ quartile & Max \\
\hline
\hline
z &  0.063   &    0.195     &  0.267     &  0.347     &  0.460\\
\hline
log Mass [M$_{\odot}$] & 8.61   &     9.86   &     10.10   &     10.50    &    11.50\\
\hline
log SFR (\ha) [M$_{\odot}$ yr$^{-1}$]& -0.83  &     0.42  &     0.85    &    1.15   &     2.32\\
\hline
log SFR(\lir) [M$_{\odot}$ yr$^{-1}$] &-1.03  &     0.45 &      0.78  &     0.99    &    1.98\\
\hline
log SSFR [yr$^{-1}$] & -10.80  &     -10.10 &      -9.65    &   -9.36    &   -8.31 \\
\hline
Z [log (O/H) + 12] &  8.10   &     8.64   &     8.76   &     8.88   &     9.17\\
\hline
\ha~ EW [$\AA{}$]&  4.35  &      19.60     &   31.00   &     46.00  &      204.00\\
\hline
\hb~ EW  [$\AA{}$]& 0.00 &        2.35 &       4.82 &        7.36 &       28.30
 \\
\hline
\end{tabular}
\caption{Properties of the 474 sample galaxies, indicated by their minimum and maximum value, as well as the median, the 1$^{st}$ and the 3$^{rd}$ quartile.}
\label{prop}
\end{table*}

\section{SED fitting: infrared luminosities and masses}

\subsection{Infrared luminosities}

To derive the L$_{IR}$ value of each source we have performed a SED-fitting using the  \textit{Le Phare} code (\citealt{Arnouts2001}, \citealt{Ilbert2006}), which separately fits the stellar part of the spectrum with a stellar library and the IR part (from 7 $\mu$m) with IR libraries.  To perform the SED-fitting we used 4 different infrared libraries (\citealt{Dale2001}, \citealt{CE2001} and Lagache cold and Lagache SB from \citealt{Lagache2004}) and set them as a free parameter, $i.e.$, for each source, the code chooses the best fitting template of all the four libraries. This helps producing more accurate fitting to the data than choosing only one IR library. The code first performs a template-fitting procedure based on a simple $\chi^{2}$ minimization; then it integrates the resulting best-fit spectra from 8 up to 1000 $\mu$m and directly gives as an output the L$_{IR}$ value. We fixed the redshift to the spectroscopic redshift ($z_s$)  of each source and made use of multi-wavelength data, including IRAC, Spitzer 24  $\mu$m and recent PEP Herschel data at 100 and 160 $\mu$m. The inclusion of these new data at large wavelengths, which  sample the critical far-infrared peak of nearby galaxies, allows us to derive the L$_{IR}$ without associated uncertainties due to extrapolation of the IR peak (see \citealt{Elbaz2010}, \citealt{Nordon2012}). We estimate the $1\sigma$ uncertainty of the L$_{IR}$ as half the difference between L$_{IR, sup}$ and  L$_{IR, inf}$, where  L$_{IR, sup}$ and  L$_{IR, inf}$ are the L$_{IR}$ values for $\Delta\chi^2 = 1$. In  Fig. 1 we show some examples of our SED-fittings. Black circles are the observed data, while the best SED fitting are the blue and the red lines for the stellar and starburst component, respectively. Notice the importance of the PEP data at 100 and 160 $\mu$m to accurately sample the IR peak of the spectra emission.

\begin{figure}
\centering
\includegraphics[angle=0,
width=0.5\textwidth]{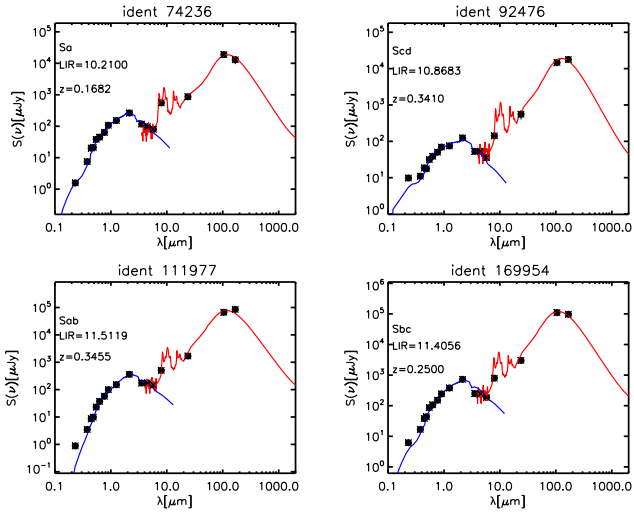}
\caption{Example of SED-fitting for 4 sources. The
observed SED of each source (black filled circles) is shown with the
corresponding best-fit solution (blue solid line for the stellar part, red solid line for the IR), as well as the morphological type (see Sect. \ref{sect:morph}), the $z_{s}$ and the derived L$_{IR}$.}
\end{figure}

\subsection{Stellar Masses}
\label{SED-mass}

The galaxy stellar masses have been derived as explained in \cite{Dominguez2011} (see also \citealt{Ilbert2010}). We have used the  $\textit{Lephare}$ code to fit our multi-wavelength data (from NUV to 5.8 $\mu$m) with a set of SED templates from \citet{Maraston2005} with star formation histories exponentially declining with time as $SFR \propto e^{- t/\tau}$. We used 9 different values of $\tau$ (0.1, 0.3, 1.0, 2.0, 3.0, 5.0, 10.0, 15.0 and 30.0 Gyr) with 221 steps in age. The metallicity assumed is solar. Dust extinction from  SED-fitting was applied to each galaxy using the \citet{Calzetti2000} extinction law. In this paper we allow a maximum $E(B-V)$ value of 1.4. This value was chosen as the highest $E(B-V)$ that the code was able to handle, given the large grid of parameters used in the SED-fitting, and is significantly larger than the highest $E(B-V)$ value that we derive for our sample ($E(B-V)_{max}$=0.85, see Sect. \ref{SFR-ha}). We imposed to the derived age of the galaxies to be less than the age of the Universe at that redshift and  greater than $10^{8}$ years. The latter requirement avoids having best fit SEDs with unrealistically high specific star formation rates, $SSFR=SFR/M$.  The \citet{Maraston2005} models include an accurate treatment of the thermally pulsing asymptotic giant branch (TP-AGB) phase, which has a high impact on modeling the templates at ages in the range $0.3 \lesssim  t \lesssim 2$ Gyr, where the fuel consumption in this phase is maximum, specially for the near-IR part; although their accurateness is still debated by a number of recent observational studies (\citealt{Kriek2010}, \citealt{Melbourne2012}).

\section{SFR from H${\alpha}$ emission line}
\label{SFR-ha}
In this paper we will derive the SFR from the H${\alpha}$ emission line luminosity corrected for extinction [SFR(\halpha) hereafter]. We will use the conversion factor derived by \cite{Kennicutt1994} and \cite{Madau1998} to obtain our SFR from the H${\alpha}$ emission line:\newline
\begin{equation}
SFR(M_{\odot} year ^{-1})= 7.9 \times 10^{-42} L (H_{\alpha})(erg~s^{-1})
\label{Eq:SFR-ha}
\end{equation}

 H${\alpha}$ emission line fluxes were measured using the automatic routine Platefit VIMOS \citep{Lamareille2009}. After removing a stellar component using \cite{BC2003} models, Platefit VIMOS performs a simultaneous Gaussian fit of all emission lines using a Gaussian profile. The slits in the VIMOS masks have a width of 1 arcsec, therefore an aperture correction for slit losses is applied to each source. Each zCOSMOS spectrum is convolved with the ACSI (814) filter and then this magnitude is compared with the \textit{I}-band magnitude of the GIM2D fits of \cite{Sargent2007}. The difference between the two magnitudes gives the aperture correction for each spectrum. This correction assumes that the emission line fluxes and the  \textit{I}-band continuum suffer equal slit losses.

A major problem when deriving SFR(\halpha) is the effect of dust extinction. Star formation normally takes place in dense and dusty molecular clouds regions, so a significant fraction of the emitted light from young stars is absorbed by the dust and re-emitted at IR wavelengths. A method to measure the extinction is to compare the observed ratio of the H${\alpha}$ and H$\beta$ emission lines with the theoretical value ($R_{th}=H{\alpha}/H\beta=2.86$). Then the reddening towards the nebular regions $E(B-V)$, following \cite{Calzetti2000} extinction law, can be written as:\\


%
\begin{eqnarray}\nonumber
A_{V}=[2.5 R_{V} \times log_{10}(R_{obs}/R_{th})]/(k1-k2)\\\nonumber
k1=2.659\times[-2.156+(1.509/\lambda_{1})-\\\nonumber
(0.198/\lambda_{1}^{2})+(0.011/\lambda_{1}^{3})]+R_{V}\\\nonumber
  k2=2.659\times[-1.857+(1.040/\lambda_{2})]+R_{V}\\    \nonumber          
 \lambda_{1}=0.48613~\mu m\\\nonumber
 \lambda_{2}=0.65628~\mu m\\\nonumber
 R_{V}=4.05\\\nonumber
 E(B-V)=A_{V}/4.05\\\nonumber
 \end{eqnarray}

%

Note that this A$_{V}$ and $E(B-V)$ values refer to the attenuation and reddening 
towards the nebular regions (as they are derived from the nebular emission lines), and not to the attenuation and reddening towards the stellar continuum usually found in literature and which do not necessarily take the same values (see Calzetti et al. 2000, \citealt{Wuyts2011a}).

Making use of the spectra of our sources from the 20k survey we can measure the observed ratio between the emission lines and obtain the extinction values. However, the quality of the spectra is not high enough to individually measure the  H$\beta$ line for the whole sample. To improve the S/N of the spectra we construct average spectra by adding spectra of galaxies with similar \lir~values. We constructed 6 average spectra composed of 31 galaxies each with different average \lir~ values: log \lir(L$_{\odot})$=[9.8, 10.1, 10.4, 10.7, 11.0, 11.3]. 

\begin{figure}
\centering
\includegraphics[angle=0,
width=.5\textwidth]{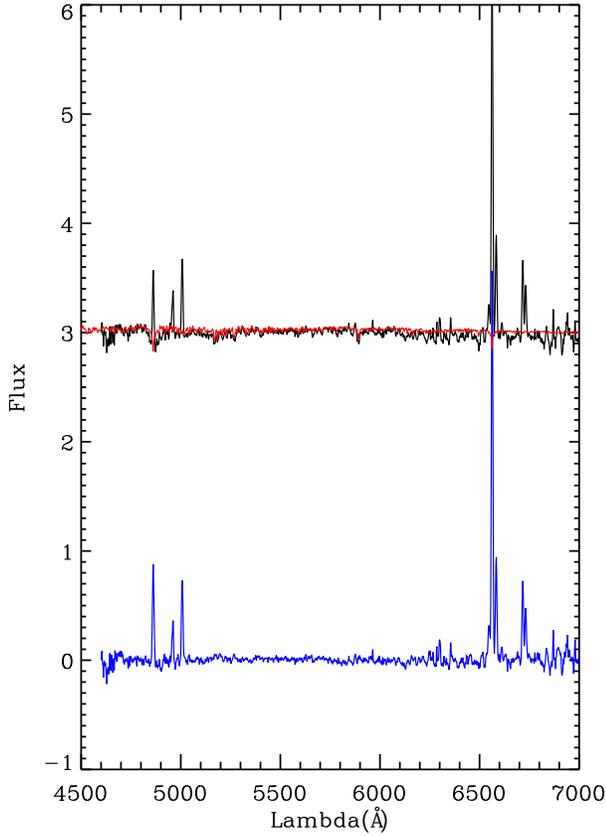}
\caption{ The black line represents the average spectrum for the lowest average \lir~before subtracting the stellar component (red), while the spectrum obtained when removing the stellar continuum is plotted in blue.}
\label{cont-subs}
\end{figure}

\begin{figure}
\centering
\includegraphics[angle=0,
width=.5\textwidth]{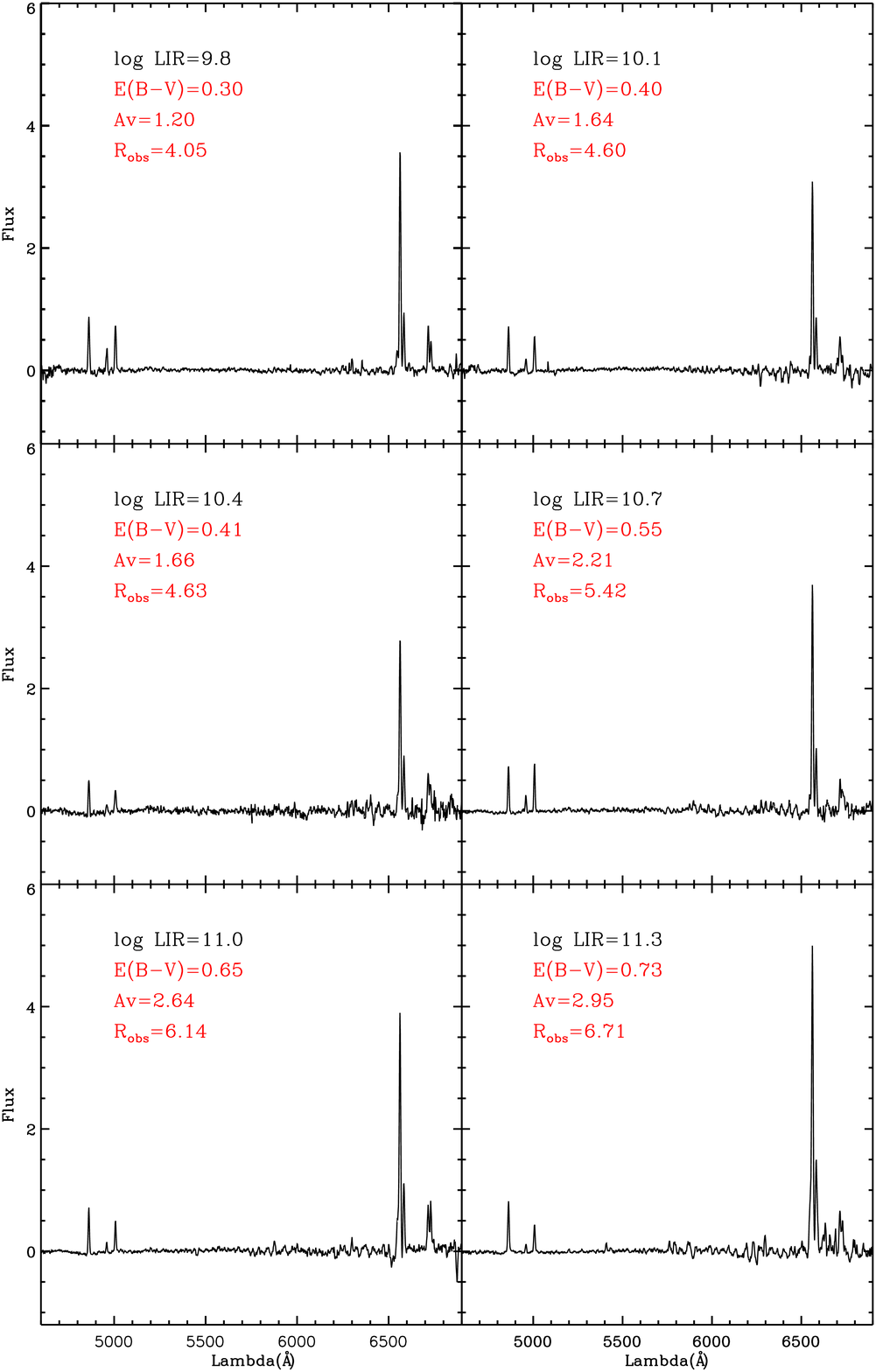}
\caption{Median spectra in 6 \lir~bins. Also shown are the values of R$_{obs}$, A$_{V}$ and $E(B-V)$.}
\label{average-spec}
\end{figure}

For each average spectrum we have measured the H${\alpha}$ and H$\beta$ fluxes using IRAF, after subtracting the stellar component making use of more than 230 \cite{BC2003} template models with different values of extinction. Fig. \ref{cont-subs} shows one of the average spectra (black line) before subtracting the stellar continuum (red line) and after subtraction of the continuum (blue line). Once the stellar component is subtracted from the observed spectra, the emission lines can be measured without contamination by the stellar absorption. Fig. \ref{average-spec} shows the 6  average spectra of our sources after performing the continuum subtraction, as well as the values of the observed ratio between  H${\alpha}$ and H$\beta$ emission lines, $A_V$ and $E(B-V)$. 

\begin{figure}
\centering
\includegraphics[angle=0,
width=.5\textwidth]{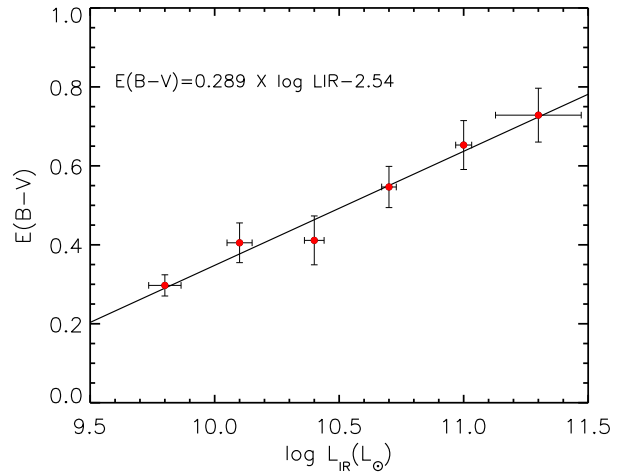}
\caption{$E(B-V)$ versus log \lir~for the 6 average spectra. The thick line is the best fit to the data, with a slope m=0.29 (see Eq. \ref{Eq:ext-lir}).}
\label{ext-lir}
\end{figure}

There is a very clear trend between the extinction  and  the \lir, as  can be seen in Fig. \ref{ext-lir}, where we plot the $E(B-V)$ value derived from the average spectra with respect to the average log \lir~of each average spectrum. The relation between dust extinction and \lir~ has also been previously studied by \cite{Wang1996}, where the authors conclude that the \hb/\ha~ratio decreases with \lir~(meaning that the dust extinction increases with \lir, in agreement with Fig. \ref{ext-lir}).  We fitted the data of Fig. \ref{ext-lir} with a straight line, obtaining a relation between log \lir~and extinction of the form:

\begin{equation}
 E(B-V)=0.29\times log L_{IR}(L_{\odot})-2.54
\label{Eq:ext-lir}
\end{equation}

The data follow the derived relation with almost no scatter. We can interpolate these data and derive an $E(B-V)$ value for each galaxy from its \lir~and correct the H${\alpha}$ flux of each galaxy making use of the \cite{Calzetti2000} extinction law:

\begin{equation}
F_{corr}(H_{\alpha})=F(H_{\alpha})\times10^{[0.4\times3.327E(B-V)]}
\end{equation}

We then derive the H${\alpha}$ luminosities using the measured spectroscopic redshift $z_{s}$  of each source, and finally we obtain the SFR(\ha) following Eq. 1.

\begin{figure}
\centering
\includegraphics[angle=0,
width=0.5\textwidth]{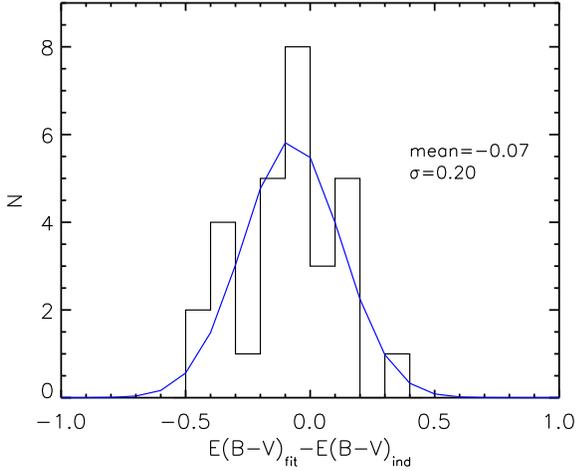}
\caption{Difference between the $E(B-V)$ value derived from the linear relation from Eq. \ref{Eq:ext-lir} and the $E(B-V)$ value measured for each source for a control sample of high S/N galaxies in the continuum both in the  H${\alpha}$ and  H$\beta$ spectral ranges. The $\Delta($E(B-V)$)$ values are consistent with a gaussian distribution with $\sigma = 0.20$. We assume this dispersion to be the error in $E(B-V)$ that we introduce when using an average extinction instead of an individual extinction.}
\label{delta-ebv}
\end{figure}

Our main source of SFR(\ha) error is likely to be the dust extinction correction. To test the effect of using a linear relation between the extinction and log \lir~instead of using single values of extinction derived for each galaxy, we have visually inspected the spectra of our sources and we have selected a small sub-sample of 29 sources with high S/N continuum in both the   H${\alpha}$ (S/N$_{c} > 2$) and  H$\beta$ (S/N$_{c} > 5$) spectral ranges, so that we can rely on the measured value of the  H${\alpha}$/H$\beta$ ratio to derive $E(B-V)$ for each source ($R_{obs, ind}$). In Fig. \ref{delta-ebv} we show the difference between the $E(B-V)$ value obtained through the linear relation between log \lir~and $E(B-V)$ and  the $E(B-V)$ value measured for each source for this control sample of 29 galaxies. The dispersion is $\sigma=0.20$, with a small, not significant offset ($\langle\Delta($E(B-V)$\rangle=-0.07$). This small offset confirms that, when using the linear relation between extinction and log \lir~instead of using individually measured extinctions, we are not introducing a systematic bias but only increase the dispersion of the data. We interpret the value $\sigma=0.20$ as the $E(B-V)$ uncertainty that we introduce in the  SFR(H${\alpha}$) derivation.

\section{SFR comparison}

In this section we will compare the SFR(\halpha) with the SFR derived from the \lir. We have converted our L$_{IR}$ values into SFR making use of the K98 relation, which simply multiplies the L$_{IR}$ by a constant:

\begin{equation}
SFR(M_{\odot} year ^{-1})= 4.5 \times 10^{-44}L_{IR}(erg~ s^{-1})
\label{Eq:K98}
\end{equation}

\begin{figure*}
\centering
\includegraphics[angle=0,
width=1.0\textwidth]{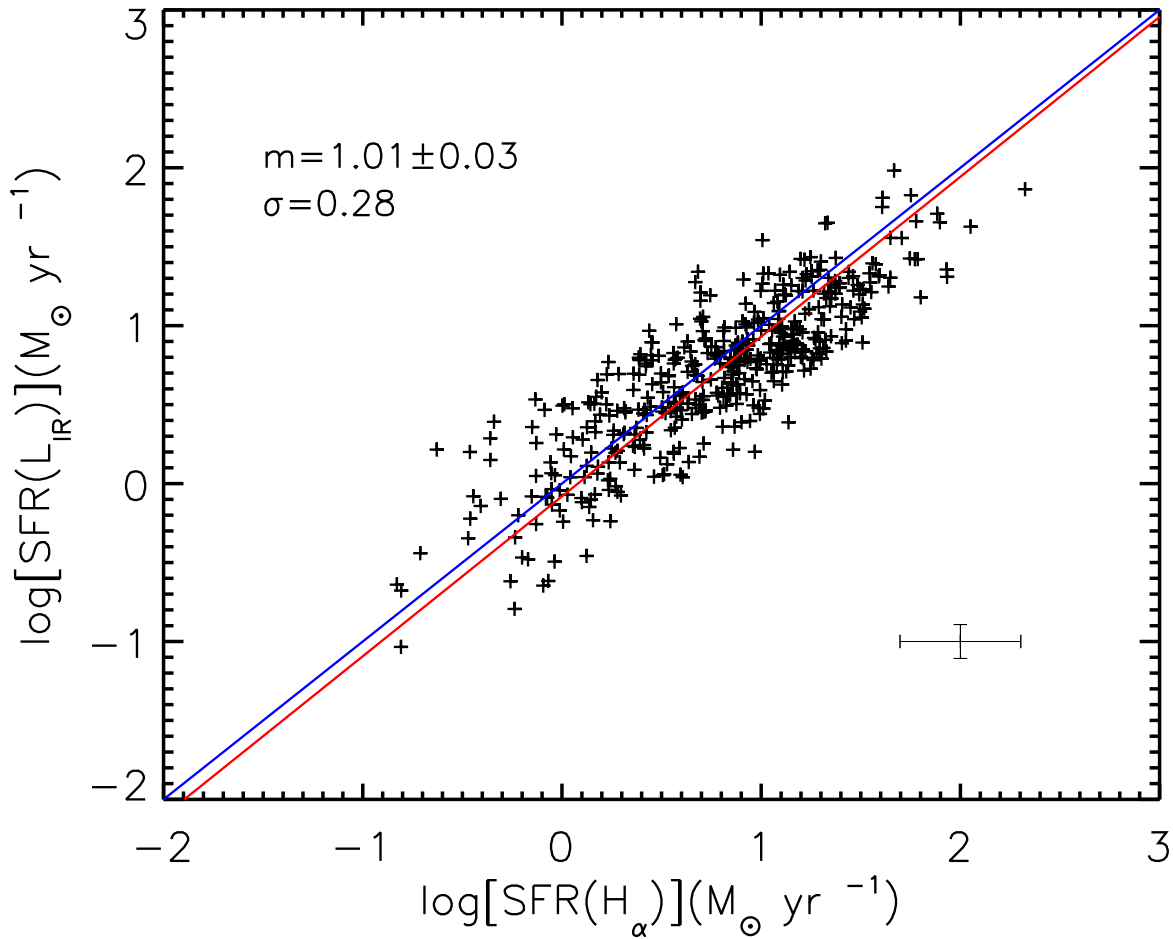}
\caption{SFR(L$_{IR}$) vs SFR(H${\alpha}$).The blue line represents the one to one relation, while the red line is the best-fit to our data. Also shown in the plot the slope of the best fit ($m=1.01 \pm 0.03$) and the dispersion of the relation ($\sigma=0.28$). The median error on log SFR(L$_{IR}$) and log SFR(H${\alpha}$) is shown towards the lower right of the plot. }
\label{SFR-SFR}
\end{figure*}

In Fig. \ref{SFR-SFR} we compare the SFR(\halpha) with the SFR(\lir). The blue line is the one to one relation, while the red line is the best fit to our data. For the clarity of the figure we do not include the errors for each source, but show the median error on log SFR(L$_{IR}$) and log SFR(H${\alpha}$) at the right hand of the plot. We have performed a linear fit taking into account the errors in both SFR(L$_{IR}$) and in SFR(H${\alpha}$). The latter include also the uncertainty introduced when using median extinctions (as explained in Sect. \ref{SFR-ha}). The slope of our relation is $m=1.01 \pm 0.03$. Also the normalizations of the two relations are well consistent with each other with a very small offset of $\sim -0.08 \pm 0.03$ dex. Therefore, there is an excellent agreement between the two SFR indicators.
 It is interesting to notice the relatively small dispersion of the data ($\sigma=0.28$), comparable with the uncertainties derived for the SFR(\halpha) (with a mean error $\sim 0.30$). This is mainly due to the large uncertainties introduced in the  SFR(\ha) derivation when using a linear relation between log \lir~ and the dust extinction values instead of using the $E(B-V)$ values for each galaxy. Considering the wide range of SFRs studied (about two orders of magnitude) and the systematic errors that could in principle affect our sample, we confirm that the two SFR estimates are in an excellent agreement, even if both of them are based on very simple recipes that do not take into account the galaxies intrinsic properties, but are a mere transformation from luminosity to SFR. 

Our results are consistent with the previous work by \cite{Kewley2002}, where the authors compare SFRs from \ha~luminosity and \lir~for a local sample of galaxies from the Nearby Field Galaxy Survey. After reddening correction, they derive a slope of $1.07 \pm  0.03$ and  a normalization constant of $-0.04 \pm 0.02$, in a very good agreement with the values that we have obtained.

It should be mentioned that, since the dust-corrected SFR(\ha) is, in the absence of 
measurement uncertainties, per definition, the total SFR,  while the SFR(\lir) is 
only the obscured part of the total, the good one-to-one correspondence between SFR(\ha) and 
SFR(\lir) implies that the unobscured SFR is much lower than the obscured SFR. This is true since our sample is PEP-detected and is in agreement with the result  of \cite{Caputi2008}, where the authors show that the SFR form the UV and the total SFR (IR+UV) differ by a factor  $\sim$ 10 on average for a 24$\mu$m selected sample from the  zCOSMOS-bright 10k survey. When studying a random, not IR detected sample of galaxies, the emission 
from young stars coming out unobscured in the UV may be significant, in which case the SFR(\lir) would not be in such a good agreement with the total SFR(\ha).

We have considered the possible bias in our results which might be induced by a selection of the  sample in the far-IR bands. For the sources with a detected H${\alpha}$ flux, but not detected in the far-IR, we have derived the SFR from the H${\alpha}$ luminosity and then transformed this into  L$_{IR}$ assuming that the K98 relation is correct (i.e. inverting Eq. \ref{Eq:K98}). There is a quite robust and well defined correlation between L$_{IR}$ and 100 and 160  $\mu$m fluxes for a given redshift. Therefore, knowing the redshift and the L$_{IR}$,  we have calculated the approximate PEP fluxes that these sources should have if they would follow the K98 relation. $99\%$ of the sources detected in H${\alpha}$ but not detected in the far-IR  have estimated PEP fluxes lower than the PEP detection limit, $i.e.$ they are consistent with following the K98 relation and being undetected in the far-IR.

 We have also studied a possible aperture correction effect. The IR fluxes at 100 and 160 $\mu$m from Herschel are integrated fluxes over the whole galaxy due to the large PSF. Instead, as explained in Sect. 4, the  H${\alpha}$ is measured with a slit of 1 arcsec width and then aperture corrected. This correction assumes that the radial H${\alpha}$ flux distribution is the same as the radial distribution of the continuum in the I(814) ACS for each galaxy. However we know that galaxies have internal structures, specially affecting the star forming regions where the H${\alpha}$ emission line is observed. Larger galaxies should be more affected by aperture correction as the observed H${\alpha}$ flux is measured in a smaller part of the galaxy, while for small galaxies almost all the light enters into the slit.  We have tested the aperture correction effect by comparing the difference between SFR(L$_{IR}$) and SFR(H${\alpha}$) with the angular sizes of the galaxies. We did not see any systematic effect, $i.e.$ the $\Delta$(SFR) does not depend on the galaxy sizes. We therefore conclude that the aperture correction does not systematically affect the slope of the \SFRSFR~relation.

Finally, we remind that our sample misses about $25\%$ objects with PACS detection but \ha~ $S/N<3$. These may include $\textquoteleft$aging' objects with low SSFR, extending trends discussed in Sect. 6.5, as well as some of the most dust obscured systems.

\section{\SFRSFR~relation dependences}

We have seen that both SFRs estimates agree very well with each other for the sample of galaxies as a whole. In this section we aim to detect which intrinsic properties of the galaxies could affect the SFR(\halpha)-SFR(\lir) relation.

\subsection{Redshift}

\begin{figure}
\centering
\includegraphics[angle=0,
width=0.5\textwidth]{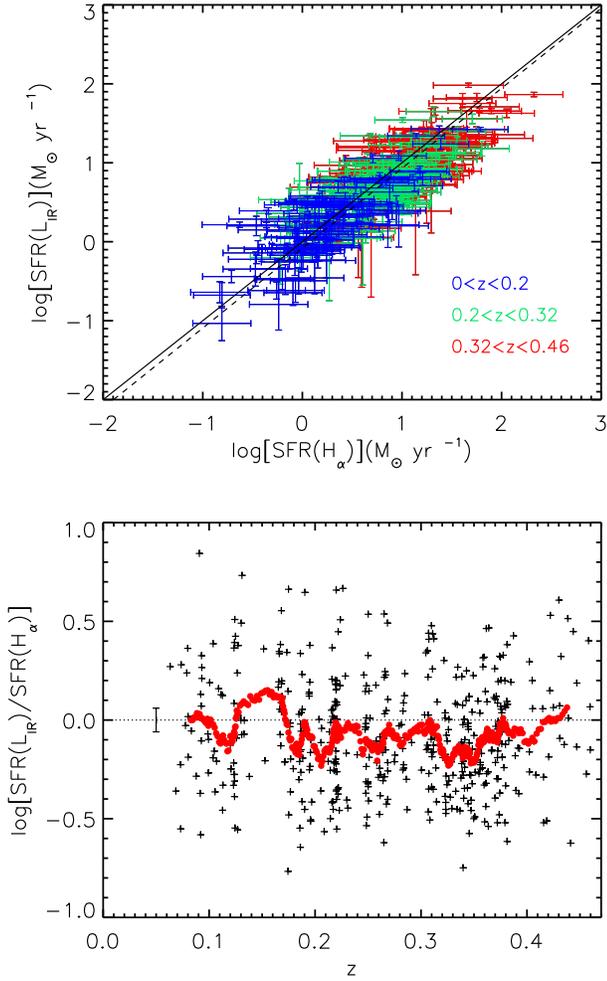}
\caption{Upper panel: log SFR(L$_{IR}$) vs log SFR(H${\alpha}$) for different redshift bins. Dashed and continuous lines represent the best fit to our data and the one to one relation, respectively. Lower panel: $\Delta(SFR)=$log SFR(L$_{IR}$)- log SFR(H${\alpha}$) versus $z$. Red dots are the running mean of $\Delta(SFR)$ every 20 values of z. Also shown at the left hand of the plot is the typical mean error of the running mean.}
\label{z}
\end{figure}

In Fig. \ref{z} (upper panel) we show  log SFR(L$_{IR}$) vs log SFR(H${\alpha}$) for different redshifts, where the color code is explained in the plot. To better appreciate the effect of $z$, we plot in the lower panel the ratio between  log SFR(L$_{IR}$) and log SFR(H${\alpha}$) versus $z$. The red dots are the running mean of the $\Delta$(SFR)=log SFR(L$_{IR}$)-log SFR(H${\alpha}$) every 20 values of $z$. The error bar on the left hand of the plot represents the typical mean error of the running mean ($\sigma_{rm,~z}=0.06$). There seems to be no significant dependence of the  relation on the redshift, but a random scatter around zero. We conclude that the \SFRSFR~relation is not highly affected by the redshift of the galaxies, at least up to $z=0.46$, which is the maximum considered $z$ of our sample.

\subsection{Mass}

\begin{figure}
\centering
\includegraphics[angle=0,
width=0.5\textwidth]{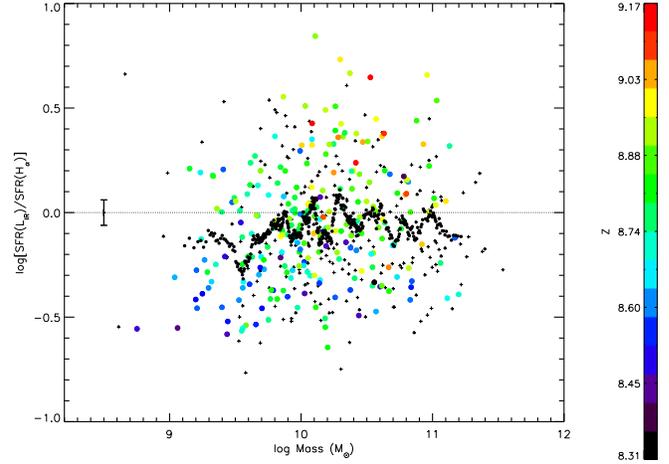}
\caption{log SFR(H${\alpha}$)-log SFR(L$_{IR}$) versus log Mass. The  black dots are the running mean of $\Delta(SFR)$ every 20 values of stellar mass, while the colors represent different metallicity values. Also shown at the left hand of the plot is the typical mean error of the running mean.}
\label{m}
\end{figure}

  In Fig. \ref{m} we show the ratio between log SFR(L$_{IR}$) and log SFR(H${\alpha}$) versus the stellar mass. The black dots are the running mean of the $\Delta$(SFR)=log SFR(L$_{IR}$)-log SFR(H${\alpha}$) every 20 values of mass and the error bar on the left hand of the plot represents the typical mean error of the running mean ($\sigma_{rm,~M}=0.06$). In this case, there is not only a random scatter around zero, but there is a slight trend with mass, in the sense that for low mass sources the SFR(L$_{IR}$) is lower than the SFR(H${\alpha}$). We have performed a Kolmogorov-Smirnov test to the distribution of the ratio between the two SFR indicators for the low mass (log M $<$ 10) and high mass (log Mass $>$ 10) galaxies, finding that the probability $P$ that the two distributions come from the same population is $P=0.037$, meaning that the behavior for the low and high mass galaxies is different at $~2.1\sigma$. 
%

We should remind that our SFR(H${\alpha}$) is also affected by errors and systematic uncertainties on its derivation. \cite{Brinchmann2004} (B04 hereafter) have shown that, differently from what is assumed in Eq. \ref{Eq:K98}, the ratio of observed H${\alpha}$ luminosity to SFR, $i.e.$ the efficiency $\eta=L_{H\alpha}/SFR$, is not constant, but depends on the mass of the galaxy, with higher values for lower mass systems. Therefore, the mass dependence that we observe could be due to the fact that we are over-estimating the SFR(H${\alpha}$) for low mass systems.

To address this issue, we have re-calculated our SFR(\halpha) making use of the B04 recipes. In this work, the authors derive different $\eta$ values for different mass ranges.  Even using mass-dependent efficiencies, the disagreement   between the two SFR indicators at lower masses is still present, while the slope of the \SFRSFR~relation remains consistent with the one to one relation ($m=0.95\pm0.03$). As a further improvement, in B04 the authors also derive a mass dependent extinction value for the \ha~emission line. We have also calculated the SFR values using both the mass dependent efficiency and the mass dependent extinction from B04 (instead of using the extinction values derived from Eq. \ref{Eq:ext-lir}). When including the extinction values derived from B04, the disagreement at lower masses is slightly reduced, but now the values of the SFR(\lir) are significantly higher than the SFR(\ha) at higher masses. Moreover, the slope of the \SFRSFR~relation derived using the B04 extinction values is significantly larger than 1 ($m=1.26 \pm 0.05$). This is mainly due to the fact that the B04 extinction values at high \lir~are lower than the extinction values that we derive from Eq. \ref{Eq:ext-lir}.

\subsection{Metallicity}

\begin{figure}
\centering
\includegraphics[angle=0,
width=0.5\textwidth]{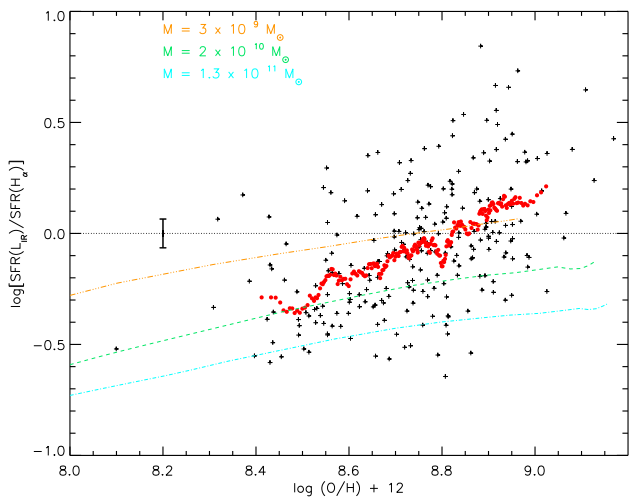}
\caption{log SFR(\lir)-log (\ha) vs metallicity. Red dots are the running mean of $\Delta$(SFR) every 20 values of $Z$. The colored lines represent the result for the chemo-spectrophotometric model for three spiral 
galaxies of different masses  (see Sect. \ref{model}).}
\label{met}
\end{figure}

A physical parameter related to the mass is the metallicity. Metallicity and mass are correlated by the well known  mass-metallicity relation \citep{Tremonti2004}, with more massive galaxies being more metal-rich, at least up to  M$\sim 10 $M$_{\odot}$. Above this mass, metallicity still increases with mass, but the slope of the relation is significantly flatter than at lower masses.  Therefore, the difference between SFR(H${\alpha}$) and SFR(L$_{IR}$) could also be due to the different metal content of the galaxies. To test this effect we have made use of the COSMOS metallicity catalog by Nair et al. (in preparation), using the \cite{Denicolo2002} method to estimate metallicity from [NII] and \ha~emission lines. In Fig. \ref{m} we have plotted our sources with different colors for different metallicity values (black crosses represent galaxies for which the metallicity measurement is not reliable, 48 $\%$ of the sample). It can be seen that the lower values of the SFR(\lir) at low masses is mainly driven by a small number of very metal-poor galaxies, while most of the most metal-rich galaxies lie above the zero point.

To better appreciate the dependence of the two SFR indicators from metallicity,  in Fig. \ref{met} we show the ratio between the two SFR indicators versus the metallicity. The red dots are the running mean of $\Delta$(SFR)=log SFR(L$_{IR}$)-log SFR(H${\alpha}$) every 20 values of $Z$. Also shown in the left hand of the plot is the typical mean error of the running mean ($\sigma_{rm,~Z}=0.07$). It can be observed that for metal-poor galaxies the SFR(\lir) values are lower than the SFR(\halpha), while for metal-rich galaxies the SFR(\lir) values are larger. The average difference between the two SFR indicators varies by $\sim 0.6$ dex when moving from the metal-poor to the metal-rich galaxies, meaning that the metallicity has a very important effect when deriving the SFR. This is consistent with the fact that more metal-rich galaxies usually have higher dust to gas ratios (\citealt{Baugh2005}, \citealt{Schurer2009}), meaning that more metal-rich galaxies are  more efficient in absorbing and re-emitting  the light emitted by young stars at IR wavelengths (at least until the optical depth in the UV goes over $\tau>1$). Therefore, for the same SFR values, a metal-rich galaxy emits more in the IR; thus the SFR derived from the \lir~is over-estimated with respect to the SFR from the \ha~luminosity. We remind that the K98 relation is calibrated for solar metallicity and we note that for that value (12+log(O/H)= 8.69, \citealt{Asplund2009}) the agreement between the two SFR indicators is very good.

 It could be argued that the observed trend with metallicity could be also affected by using the linear relation between log \lir~and $E(B-V)$ for all the galaxies. For metal-poor galaxies, the dust extinction correction that we apply could be larger than the actual one; thus the SFR(\ha) that we derive would be biased high, leading to the low values of the ratio between the two SFR indicators at low metallicities.  However, even when using the B04 recipes to derive the SFR, which include both a mass-dependent efficiency and a mass-dependent dust extinction, the ratio between the two SFR estimates was still very different for metal-poor and metal-rich galaxies. We eliminated the scatter due to the metallicity dependence on the SFR comparison, finding a reduction of the dispersion of 14$\%$ (from $\sigma=0.28$ to $\sigma=0.24$). We conclude that metallicity is a key parameter when deriving the SFR and suggest that the K98 relation cannot be applied blindly when considering very  metal-poor/rich galaxies.

\subsection{Comparison with a model for dust evolution in spiral galaxies}
\label{model}

A suite of chemical evolution models for spiral galaxies able to
reproduce the observed evolution of the mass-metallicity relation  has
been presented in Calura et al. (2009).  In that work, three spiral
models, representing galaxies of different masses, have been
designed in order to reproduce a few basic scaling relations
of disk galaxies across a stellar
mass range spanning from $10^9 \, M_{\odot}$ to $10^{11} \, M_{\odot}$.
In such models, it was assumed that the baryonic mass of any spiral 
galaxy is dominated by
a thin disk of stars and gas in analogy with the Milky Way.
The disk is approximated by several independent rings, 2 kpc wide, 
without exchange of matter between them.
The timescale for disk formation is assumed to increase with the 
galactocentric distance,
according to the ``inside-out'' scenario \citep{Matteucci1989}.
The models for spiral disks include dust production, mostly from 
core-collapse supernovae and intermediate mass stars,
restoring significant amounts of dust grains during the asymptotic giant 
branch phase,
as well as dust destruction in supernova shocks and dust accretion.
The prescriptions for dust production are the same as in \cite{Calura2008}, where a chemical evolution model for dust was presented, able to 
account for the observed dust budget in the
Milky Way disk, in local ellipticals and dwarf galaxies.
In a subsequent work, \cite{Schurer2009} combined the predicted dust 
masses and the star formation
history of the MW galaxy
with a spectrophotometric evolution code that includes dust reprocessing 
(GRASIL).

The relation between log  [SFR(\lir)/SFR(\ha)] and metallicity 
observed in
spiral galaxies is compared
with the results from chemo-spectrophotometric models of 
three spiral galaxies of different masses in Fig. \ref{met}.
In the model, the SFR(\lir) quantity is calculated according to Eq. \ref{Eq:K98},
where the total IR luminosity $L_{IR}$ of the spiral galaxy has been 
calculated by
integrating the SED in the range  $8 - 1000\mu$m.
In Fig. \ref{met}, the metallicity (expressed as log (O/H) +12) is a direct 
output of the chemical evolution models
of Calura et al. (2009). Here, the quantity SFR(\ha) is
the intrinsic SFR $\psi$, calculated by means of the Schmidt-Kennicutt law
\begin{equation}
\psi \propto \sigma^{k}
\end{equation}
where $\sigma$ is the gas density per unit surface and $k\sim 1.5$.
In agreement with what is observed in real galaxies,
in the three models the ratio log [SFR(\lir)/SFR(\ha)] grows with 
metallicity.
This is due to the fact that, as galaxies evolve, their interstellar 
metallicity rises
and their star formation rate decreases, accompanied by a milder 
decrease of the FIR luminosity.
In the model, the smaller decrease of the FIR luminosity with respect to 
the SFR
reflects the fact that the FIR luminosity is only to a first order 
approximation
proportional to the SFR; in fact, significant contributors to dust 
production and heating are
also intermediate mass stars, whose total mass within a galaxy reflects 
the integrated star formation history.
Our models predict a larger log [SFR(\lir)/SFR(\ha)] value
at a fixed metallicity for lower mass galaxies.
It is also worth noting that the observed increase of the 
log [SFR(\lir)/SFR(\ha)]- metallicity
relation is stronger than the model results, with a substantial number of  the 
observed systems presenting
log [SFR(\lir)/SFR(\ha)]$>0.1$, not reproduced by our models.
The majority of these systems have metallicities $12+log (O/H)> 8.5$, but do
not present particularly high SFR or stellar mass values.
There are two possible reasons for this discrepancy between the model 
results and the
observed galaxies.

One possibility is that the star formation history of the three spiral 
models
does not represent the true star formation history of the observed galaxies; a significant 
number of real galaxies may
have more complicated star formation histories, for instance 
characterized by episodic starbursts,
as opposed to our models, characterized by smooth star formation histories.
Another possibility is that, for these relatively high metallicity systems,
the corrected SFR(\ha) values may represent lower limits to the 
intrinsic SFR values.
It could be that, the larger the metallicity and the dust content of the 
galaxy,
the more likely are the star forming regions to become thick to \ha~
radiation.
One consequence may be that particularly optically thick regions may be 
missed, and that the total
SFR rate determined from \ha~ emission may be underestimated.
At present, it is very difficult to assess which one of the two 
possibilities may be the main cause of our discrepancy.
A useful test in this regard would imply the use of spectro-photometric 
models including the effects of dust extinction
on nebular emission lines ($e.g.$ \citealt{Panuzzo2003}).
The combined use of such models with chemical evolution models including 
dust evolution will
be an interesting subject for future work and will likely shed more 
light on the main subject of this paper.

\subsection{SSFR}

\begin{figure}
\centering
\includegraphics[angle=0,
width=0.5\textwidth]{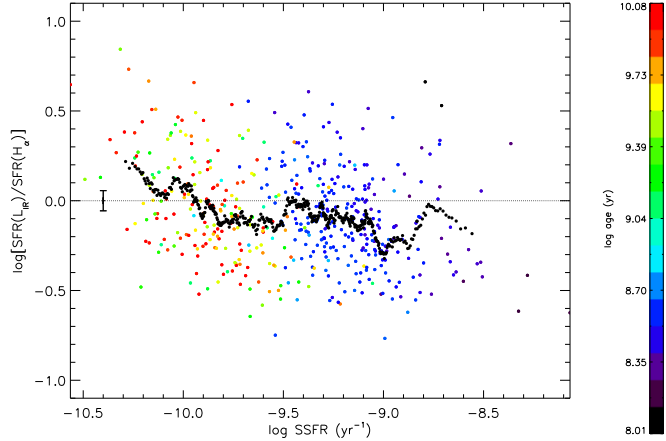}
\caption{log SFR(L$_{IR}$)-log SFR(H${\alpha}$) vs SSFR. Colors represent the galaxy stellar ages as derived from the SED-fitting. The black dots are the running mean each 20 values of SSFR and the error bar at the left hand of the plot is the typical mean error of the running mean.}
\label{ssfr}
\end{figure}

The SSFR is a very interesting parameter as it does not only account for the number of stars being formed, but also takes into account the efficiency of a galaxy of a given mass to form stars with respect to its total mass content. It gives information about the star formation activity of a galaxy, allowing us to divide them into star-forming, active or intermediate galaxies. It is reasonable to consider that galaxies with different star-formation modes may show different luminosity-SFR relations. In Fig. \ref{ssfr} we show a similar plot to Fig. \ref{m} where now the considered parameter is the SSFR and the colors represent different ages of the galaxies derived by means of the SED-fitting (calculated when deriving the galaxy stellar masses as explained in Sec. \ref{SED-mass}). Also shown is the typical mean error of the running mean  ($\sigma_{rm, SSFR}=0.06$). The SFR value that we use in this plot is the mean value of SFR(\ha) and SFR(\lir). This value has been chosen to avoid a systematic trend with one of the two SFR indicators, which are present in the \textit{y} axis. As expected, all of our galaxies have high values of SSFR and would be classified as star-forming galaxies following different criteria recently found in literature (see $e.g.$ \citealt{Dominguez2011}, \citealt{Pozzetti2010}).

We observe values of the SFR(\lir) larger than the SFR(\ha) by $\sim$ 0.2 dex  for galaxies with small values of SSFR ($<$ 10.0 [yr$^{-1}$]). A possible explanation for these higher values of SFR(\lir) at low SSFRs is the fact that the old stellar population may significantly contribute to the \lir~of a galaxy  as already mentioned by K98 (see also a recent work by \citealt{daCunha2011}). For the galaxies with low SSFR, the heating by stars older than 10$^{7}$ yr in the diffuse inter stellar medium can be comparable to the dust luminosity from the absorption of light from young stars; thus the value of the SFR(\lir) is overestimated as it includes emission from the old stellar population and not only from the recently formed stars. In fact, as it can be seen in Fig. \ref{ssfr}, the oldest galaxies (redder colors) populate the low SSFR region of the plot, while the galaxies with the youngest stellar populations (bluer colors) have always larger SSFR inferred values.
Another possible explanaition for the observed trend between the two SFR indicators with SSFR  could be due to spikes in recent SFR. A galaxy undergoing a current burst (high SSFR) will have a high \ha~ flux compared with its \lir~ and the   log [SFR(\lir)/SFR(\ha)] will take negative values, with  the opposite trend happening for low SSFRs.
Again, this trend is not reduced when using B04 recipes to estimate the SFR(\ha).  Finally, we would like to remark that the results from Figs. \ref{met} and \ref{ssfr} are in agreement with the fundamental metallicity relation observed by \cite{Mannucci2010}, where the authors find that a decrease in metallicity also correlates with an increase in SSFR. The SFR(\lir) seems to be higher, on average, than the SFR(\ha) both for galaxies with high metallicity or low SSFR values.

We conclude that caution must be taken when deriving the SFR from the \lir~for galaxies with low values of SSFR, as the emission from the older stellar population may overestimate the derived SFR. The opposite effect may happen for galaxies with recent satr formation bursts, $i.e.$ high SSFR values.

\subsection{Morphological type}
\label{sect:morph}

\begin{figure*}
\centering
\includegraphics[angle=0,
width=1.\textwidth]{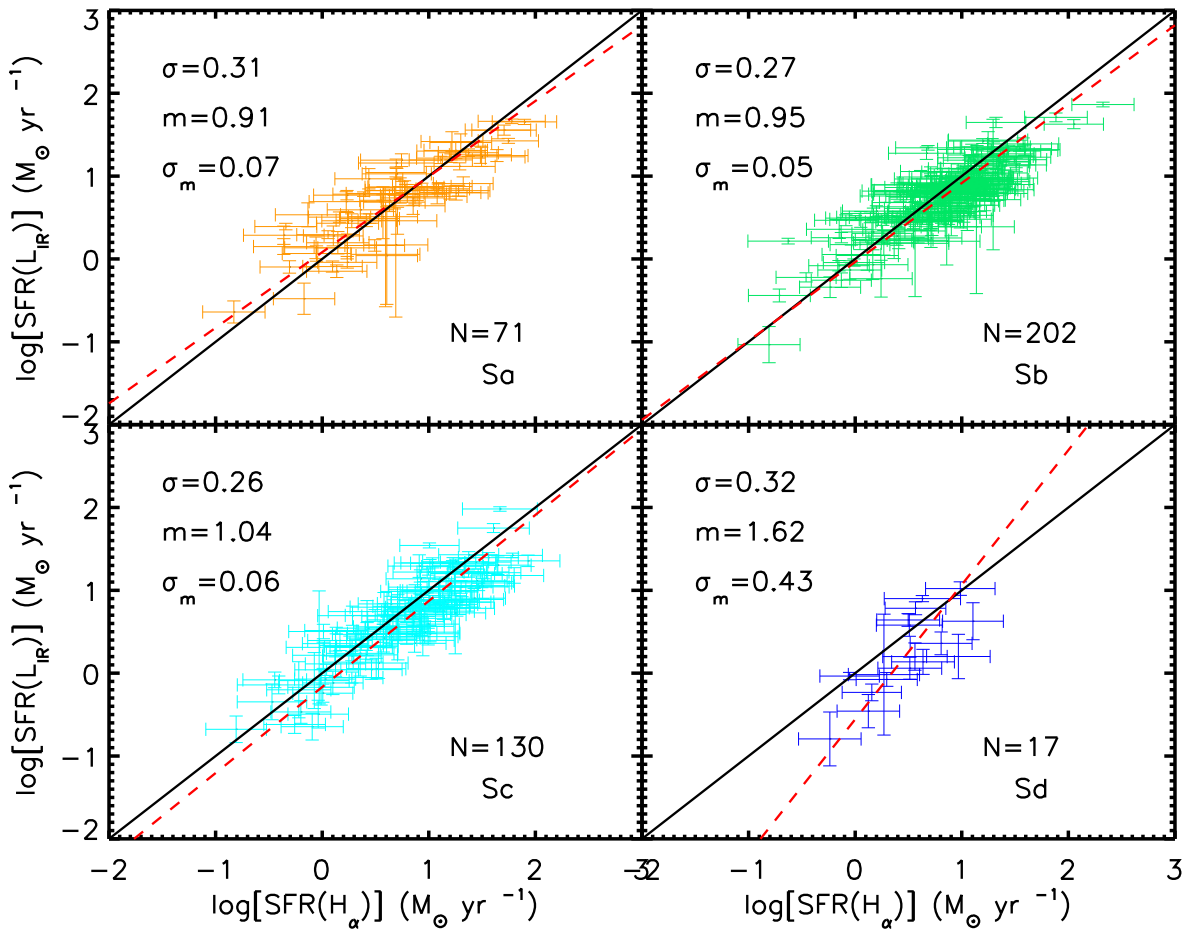}
\caption{log SFR(L$_{IR}$) vs log SFR(H${\alpha}$) for sources classified as Sa, Sb, Sc or Sd/Irr. Red dashed and continuous lines are the best fit to our data and the one to one relation, respectively. Also shown the derived values for the slope (m) and its error ($\sigma_m$), as well as the scatter of the relation ($\sigma$) and the number of sources for each morphological type.}
\label{types}
\end{figure*}

Another important property that could affect the SFR indicators is the morphological type, as different morphologies are associated with different star formation histories and geometries and therefore different SEDs and IR emission. To assess the effect of morphology on the SFR derivation we have made use of the morphological catalog by Nair et al. (in preparation), who, following a similar scheme as in \cite{Nair2010}, visually classified the galaxies belonging to the COSMOS field, dividing them into different types (Ell, S0, Sa-Sd, Irr). In Fig.\ref{types} we show the usual log SFR(L$_{IR}$) vs log SFR(H${\alpha}$) plot dividing our sample into different morphological types.

The two SFR indicators are in very good agreement with each other for the three spiral sub-samples. For each of them the \SFRSFR~ relation is consistent, in both normalization and slope, with the one-to-one relation.
For the Sd, irregular sample (Sd/Irr), the formal best-fit slope becomes much steeper ($m=1.62 \pm 0.43$), but is still consistent  with 1 at the $1.5\sigma$ level, because of the reduced statistics of this sub-sample (we have only 17 sources classified as Sd or irregular galaxies). The dependence of the SFR indicators with morphology has also been studied by \cite{Kewley2002}, finding that the ratio between SFR(\lir)  and SFR(\ha) is almost the same for early (S0-Sab) and late (Sb-Irr) type galaxies.
 If real, one possible explanation for the discrepancy for the Irr/Sd galaxies is their complex morphology. Irregular galaxies are clumpier and less homogeneous than earlier spiral galaxies, thus making it more difficult and uncertain to measure the total H${\alpha}$ flux of the galaxy due to aperture corrections. We have also investigated the possible effect of metallicity in the Sd/Irr sample, as irregular local galaxies are found to be metal poor \citep{Tolstoy2009}. However, we did not see a different metallicity distribution for the Irr/Sd sample than for the whole sample of galaxies. Despite the fact that the number of Sd/Irr objects is quite small, we observe a slight trend of a steepening of the slope of the \SFRSFR~relation when moving from early to late spiral galaxies.  For example, if we analyze together Sa and Sb galaxies we obtain a slope of $m=0.93 \pm 0.04$, $i.e.$ almost a 3$\sigma$ difference with respect to the slope for the Sc sample. This difference may be indicative of a weak dependence of the SFR indicators  with morphology. However, a more significant statistical sample would be necessary to assess if the observed steepening of the slope with morphological type is a real effect or if it is just due to the few sources considered for each morphological type, specially for the Sd/Irr sample.

\section{Summary and Conclusions}

In this paper we have empirically tested the relation between the SFR derived from the \lir~and the SFR derived from the \ha~emission line. We have studied a sample of 474 galaxies at z=0.06-0.46 with both \ha~ and IR detection from 20k zCOSMOS and  PEP Herschel at 100 and 160 $\mu$m. The L$_{IR}$ has been derived by integrating from 8 up to 1000 $\mu$m the best-fitting SED to our IR data  from IRAC, Spitzer and  Herschel. We have derived our SFR from the  H${\alpha}$ extinction corrected emission line. We have constructed 6 median spectra with different median \lir~values and we have found a very clear trend between $E(B-V)$ and log \lir. This allows us to estimate $E(B-V)$ values for each galaxy  (see Eq. \ref{Eq:ext-lir}) even if the quality of the spectra does not allow to accurately measure the \hb~ emission line for each galaxy. Our main conclusions are:

\begin{itemize}
 \item{We have compared the SFR(\lir) and SFR(\ha) finding an excellent agreement between the two SFR estimates for the bulk of the studied galaxies, with the slope of the \SFRSFR~relation $m=1.01 \pm 0.03$ and the normalization constant of $a=-0.08 \pm 0.03$. This means that the simple recipes used to convert luminosity into SFR by simply scaling relations are consistent with each other. The agreement between the two SFR estimates also implies that the  assumptions, for example on star formation histories, that were made to derive the SFR recipes (Eq. \ref{Eq:SFR-ha} and \ref{Eq:K98}) should not be very different from the properties of the studied sample. The comparison of the SFR indicators at low redshift is crucial to test the validity of the SFR estimates which are often used in high redshift studies. The main result that we obtain is that both methods are in very good agreement for the low redshift sample, allowing us to extend their validity at higher redshifts. This allows us to derive the SFR of distant galaxies when only the \lir~or the \ha~information are available, with the assurance that we are not introducing important systematic effects.}

 \item{There seems to be no dependence of the SFR indicators with redshift, at least up to $z\sim 0.46$, which is the limit of our sample.}

 \item{The stellar mass seems to have a small influence on the \SFRSFR~relation in the sense that for low mass galaxies the SFR(L$_{IR}$) values are lower than the  SFR(H${\alpha}$). However, the observed dependence of the SFR indicators with the stellar mass seems to be mainly driven by metallicity. }

\item{The metallicity, in fact, seems to be the parameter that most influences the SFR comparison, with average values of log [SFR(\lir)/SFR(\halpha)] which differ by $\sim 0.6$ dex  from  metal-poor to metal-rich galaxies. This is due to the higher efficiency in absorbing and re-emitting the light from young stars in the IR for more metal-rich galaxies. The dispersion of the \SFRSFR~relation  is reduced by $14\%$ when taking into account the scatter due to the metallicity.  We stress that caution must be taken when deriving SFR for very metal-poor/rich galaxies.}

\item{We show how the 
behavior of the observed
SFR(\lir)/SFR(\ha)-metallicity relation finds a natural 
explanation within the frame of a complete theoretical model for dust evolution in spiral galaxies.}

\item{The SFR(\lir) values are larger than the SFR(\ha) for galaxies with low SSFR values (log SSFR $<$ 10 [yr$^{-1}$]) probably due to the non negligible contribution to the IR emission from the old stellar population. Stars older than 10$^{7}$ yr may significantly contribute to the \lir~in systems with little star formation activity. The difference between the two estimates could also be affected by the effect of short and intense bursts of star formation, $i.e.$ galaxies with high SSFR values.}

 \item{When separately studying the SFRs indicators as a function of morphological types, we find that they are in excellent agreement with each other for the three spiral sub-samples. For each of them the \SFRSFR~ relation is consistent, in both normalization and slope, with the one-to-one relation. For the Sd, irregular sample (Sd/Irr), the formal best-fit slope becomes much steeper ($m=1.62 \pm 0.43$), but is still consistent  with 1 at the $1.5\sigma$ level, because of the reduced statistics of this sub-sample.}

\end{itemize}

\section{Aknowledgements}

This work was supported by an international grant from Instituto de Astrofisica de Canarias (IAC) with funding from the Research Structure Departments of Instituto Nazionale di Astrofisica (INAF). PACS has been developed by a consortium of institutes led by MPE(Germany) and including UVIE (Austria); KU Leuven, CSL, IMEC (Belgium);
CEA, LAM (France); MPIA (Germany); INAF-IFSI/OAA/OAP/OAT, LENS,
SISSA (Italy); IAC (Spain). This development has been supported by the funding
agencies BMVIT (Austria), ESA-PRODEX (Belgium), CEA/CNES (France),
DLR (Germany), ASI/INAF (Italy), and CICYT/MCYT (Spain). We thank the COSMOS/zCOSMOS teams for making the data available for the comunity and the anonymous referee for a constructive report.

\label{lastpage}

\bibliography{bibliografia}

\end{document}